\documentclass{aa}
\usepackage{natbib}
\usepackage{epsfig}
\usepackage{latexsym}   

\begin{document}
\newcommand{\mercator}{{\sc mercator}}
\newcommand{\gdor}{$\gamma$\,Dor}
\newcommand{\cd}{\,$\mathrm{d}^{-1}$}

\title{Long-term photometric monitoring with the Mercator telescope. 
Frequencies and multicolour amplitudes of $\gamma$ Doradus stars
\thanks{Based on observations collected with the Flemish 1.2-meter {\sc Mercator} Telescope at Roque de los Muchachos, La Palma} }

\author{J.~Cuypers\inst{1}  
   \and C.~Aerts\inst{2,3}
   \and P.~De Cat\inst{1} 
   \and J.~De Ridder\inst{2}
   \and K.~Goossens\inst{2}
   \and C.~Schoenaers\inst{1,4}
   \and K.~Uytterhoeven\inst{2,5,9}
   \and B.~Acke\inst{2}\fnmsep\thanks{Fellow of the Fund for Scientific Research, Flanders (FWO)}
   \and G.~Davignon\inst{2,5}
   \and J.~Debosscher\inst{2}
   \and L. ~Decin\inst{2,\star\star}   
   \and W.~De Meester\inst{2}
   \and P.~Deroo\inst{2,10}\thanks{NASA Post-doctoral fellow}  
   \and R.~Drummond\inst{2,8}
   \and K.~Kolenberg\inst{2,7}
   \and K.~Lefever\inst{2,8}
   \and G.~Raskin\inst{2,5}
   \and M.~Reyniers\inst{2,6}
   \and S.~Saesen\inst{2,\star\star}
   \and B.~Vandenbussche\inst{2}
   \and R.~Van Malderen\inst{2,6}
   \and T.~Verhoelst\inst{2,\star\star}
   \and H.~Van Winckel\inst{2}
   \and C.~Waelkens\inst{2}
}
\institute{Koninklijke Sterrenwacht van Belgi\"e, Ringlaan 3, 1180 Brussel, Belgium
\and       Instituut voor Sterrenkunde, K.U.Leuven, Celestijnenlaan 200 D, 3001 Leuven, Belgium
\and       Department of Astrophysics, Radboud University Nijmegen, 6500 GL Nijmegen, the Netherlands
\and       Oxford Astrophysics, University of Oxford, Oxford, OX1 3RH, UK
\and       \mercator\ Telescope, Calle Alvarez de Abreu 70, 38700 Santa Cruz de La Palma, Spain
\and       Koninklijk Meteorologisch Instituut, Ringlaan 3, 1180 Brussel, Belgium
\and       Institut f\"ur Astronomie, Universit\"at Wien, T\"urkenschanzstrasse 17, 1180 Wien, Austria
\and       Belgisch Instituut voor Ruimte-A\"eronomie, Ringlaan 3, 1180, Brussel, Belgium
\and       Service d'Astrophysique, IRFU/DSM/CEA Saclay, L'Orme des Merisiers, 91191 Gif-sur-Yvette Cedex, France 
\and       Jet Propulsion Laboratory, 4800 Oak Grove Drive M/S: 183-900, Pasadena, CA 91109, Uinited States
}

\offprints{J.~Cuypers}
\mail{Jan.Cuypers@oma.be}
\date{March 05, 2009}

\abstract
  { \gdor\ stars are excellent targets for asteroseismology since the gravity modes present in these stars probe the deep stellar interiors. Mode identification will improve the knowledge of these stars considerably. 
 }
  {A selected group of \gdor\ stars and some candidates were observed with the Mercator telescope to find and/or confirm the periodicities in the light variations and to derive reliable amplitude ratios in different pass bands. }
  {A frequency analysis was performed on all new data obtained in the Geneva photometric system. In order to have more reliable and accurate frequencies, the new data were combined with similar data from the literature and with Hipparcos observations. 
  A set of frequencies that minimized the the residuals in a harmonic fit was searched for while allowing means and amplitudes to vary from one observation set to another.}
  {Frequencies and amplitudes in the photometric passbands of the Geneva system are given for 21 \gdor\ stars. We report the discovery of \object{HD 74504} as a newly found  \gdor\ star.}
  {Our study provides the first extensive multicolour database for the understanding of gravity modes in F-type stars}

\authorrunning{Cuypers et al.}
\titlerunning{Photometric monitoring of \gdor\ stars} 
\maketitle

\section{Introduction: $\gamma$ Doradus stars (\gdor\ stars)} \label{intro}

As reviewed in the article by \citet{henry07} the class of \gdor\ stars has now over
60 members. These stars have spectral types of late A or F, luminosity class IV or V
and exhibit periodicity in the light variations with periods in the range of 0.3 to 3
days. Line-profile variability is present as well \citep{mathias04, decat06}.
There is no doubt that the cause of the variations is pulsation. The modes are
high-order gravity modes (g-modes), excited by a flux blocking mechanism
at the base of the convective envelope of the stars \citep{guzik00,dupret05}.

Because the g-modes probe the deep stellar interior, the \gdor\ stars are excellent
targets for asteroseismology, particularly also since they may exhibit solar-like oscillations as well \citep{michel08}.
However, mode identificiation is not simple for these
stars since the spectrum of g-modes is very dense and only a few modes seem to have
high enough amplitudes to be well observed from the ground. Therefore, observables such as
photometric amplitude ratios and line-profile variations are extremely useful to
identify the modes. By observing light variations in different passbands the
identification of the degree $\ell$ of the pulsation mode becomes possible
\citep{balona79, watson88, dupret05}.

Since this kind of observable is made available with this paper for 
21~\gdor\ stars, a large number of variable stars of this class will have their
modes identified, and parameters for the physics of stellar models, including
e.g. a value for the mixing-length parameter, can in principle be derived.
Such seismic inferences are the subject of an accompanying paper (Miglio et al, 2008, in preparation).

\section{The instrument and the observations}

The Mercator telescope is a 1.2-meter telescope located on the Roque de los
Muchachos observatory on La Palma, Spain. Since the beginning of its
scientific operations in spring 2001 it was equipped with the P7 photometer.

The P7 photometer is the refurbished photometer that has been active on the
70-cm Swiss Telescope at La Silla Chile before. It is a two-channel photometer
for quasi simultaneous 7-band measurements ($U, B_1, B, B_2, V_1, V, G$) in the Geneva photometric system
\citep{golay66, rufener88, rufener89}. The
first channel (A) is centred on the star while the second channel (B) is centred
on the sky. The filter wheel turns at 4 hertz and a chopper directs both
channels alternatively to the photomultiplier. As such, the photomultiplier
measures both beams through the seven filters four times per
second. In this way quasi-simultaneous measurements of the light in the 7
filters become possible. This makes the instrument very suitable to
observe variable stars, since information on the light variations in all filters
can be obtained very efficiently.

Furthermore an almost continuous access to the telescope during the year was
possible, so the instrument became ideal to monitor stars with periods of the
order of days and/or long beat periods of the order of months or years.
Here we report on the results of a long-term monitoring of \gdor\ stars, 
while \citet{decat07} discussed the results for a sample of variable OB stars.

Some of these stars were observed over two years (time span around 370 days), some
over three years (time span around 740 days) or four years (1110 days). Stars
observed during only one season were not considered in detail. 
Up to five observations per star were obtained for about two months per season.
This resulted in about 130 measurements per star on average, 
but there is a large spread as can be seen in Table \ref{resM}.

Photometric observations in the Geneva system are labeled with two quality indices
(from lowest quality 0 to highest quality 4), one for the magnitudes, one for the
colours \citep{rufener89}. In a first analysis we eliminated all observations labeled 0 or 1, but
we also carried out a re-analysis with the low-quality observations
included. In most cases the difference in the result was negligible. Since some
stars were only measured at high airmasses, they almost all automatically received a
low index value. Here we used those values as well, since without them the number of
observations of those stars would be too low. The errors are slightly larger in such
cases, but the results are the same. Observations of standard stars indicate
that the errors for the programme stars, which have visual magnitudes between 7 and 9, are
in the range $0.005$ to $0.010$ mag.

\section{Selection of the targets}

A selection of stars was made out of the lists by \citet{aerts98} and \citet{handler99}, of the Hipparcos
selection results and from lists of \gdor\ stars,
candidate \gdor\ or other variables \citep{koen02}. Since the start of the
observations, almost all stars of the list were comfirmed as \gdor\ stars.
For very few of them, multicolour photometry was available and in
several cases the multiperiodicity needed confirmation. The list with the 21 stars with
more than 25 observations of good quality, spread over more than one season, is given in Table \ref{resM}.
A number of stars was already (partly) analysed in a previous paper \citep{deridder04}
but for some stars the additional observations proved to be very
useful to clarify the multiperiodicity of the light variations.

Fifteen other variable A-F stars were monitored during the same period. 
Not all showed characteristics of \gdor\ stars: a few turned out to be $\delta$ Scuti stars, others need some further analysis.
The results for these stars will be published in another article.

\section{Analysis methods} 

We used standard methods for the period analysis, but since the light curves of these stars are very sinusoidal, we adopted 
as a final analysis method a multifrequency least-squares fit with sinusoidal components
\citep{schoenaers04}, where we searched for the best frequencies in large intervals around an initial solution.
This is an extremely computer intensive procedure since the sum of the residuals is a rapidly oscillating function, especially when up to six frequencies had to be considered.
To obtain the initial values we used Lomb-Scargle periodogram methods \citep{lomb76,scargle82}, 
and phase dispersion minimization methods \citep{stellingwerf78}, 
each time followed by a prewhitening for a first analysis. 
The final search was always done with a multifrequency least-squares fit. 

For initial values of the frequencies in the Mercator data we usually looked at the B or B1 data, since we
expect the highest amplitudes of the light variations in these filters (e.g. \citet{aerts04}).
It happened that the amplitude in other filters (typically V1 or G) became too low for the frequencies to be significant there. 
We re-analysed the Hipparcos data independently, but we looked only for a local minimum in the least-squares fit in the (wide) neighbourhood of the ground-based solution. 

If observations were available from other sources, we combined the Geneva colour V, Johnson V  and Hipparcos data.
We searched in the combined data for a multifrequency least-squares fit with unknown means and amplitudes for each data source to search for the best frequency (frequencies) to describe all data. We did not use statistical weights on individual observations or groups of data:
the data from the literature were of similar quality and the measurement errors about the same as for the Geneva photometry. 
The Hipparcos data often had larger errors, but since these are important for solving the alias problems, we also gave those equal weights.

Although for individual data sets only  frequencies peaking four times above the average signal-to-noise level in the periodogram are acceptable,
we relaxed this criterion for the frequencies found with the multifrequency fit technique in the combined data.
Frequency solutions of the multifrequency fit were accepted, if a corresponding extremum was found in the periodogram of each data set.
In this way we accepted frequencies with peaks 1.5 times above the noise in some of the individual data sets, as shown a posteriori. 

In several cases the combination of the data solved the alias or ambiguity problems and it gave, in general, more accurate values for the frequencies. In a few cases, it turned out that observations of the same star(s) were carried out, unintentionally, as in multisite campaigns. Here the gain in frequency determination was obvious. Including the Hipparcos data enlarged the time basis to about 5000 days for almost all stars. We did no colour transformation for the Hipparcos data, since our method takes possible amplitude differences into account. 
Since the Hipparcos data are not always as accurate as the ground-based data and since the observations sometimes have a rather odd time distribution, there was not always the expected gain in frequency accuracy. 
In most cases however, the frequency errors became of the order of $10^{-6}$\cd.

Amplitude ratios and other characteristics were computed in the different passbands of the Geneva photometric system with frequencies as close as possible to the results of the combined data set. For this purpose we searched again for the closest local minimum in the residuals of the harmonic fit in the Mercator data. In some cases there were differences between the best solution found in the Mercator data alone and this solution. However, differences were small or caused by choosing the wrong alias in the initial solution, as can be seen by comparing Table \ref{resM} and the Tables 5 to 25 (only available electronically).

\section{Results on the individual stars}

\subsection{\object{HD 277} = HIP 623}

HD 277 is an F2 star intensively observed by \citet{henry01}. They found three
frequencies and remarked more scatter around light maximum than around light minimum.

The three frequencies found by  \citet{henry01} were confirmed, although the $2-f$ alias of their third frequency is more significant in the Mercator data. The high significance of this alias is caused by the limited nightly sampling of our data, combined with the length of the period.

The third frequency as derived from the Hipparcos data (1.3437\cd) deviates from the third frequency (1.3866\cd)
found by \citet{henry01} and from our third frequency (1.38709\cd) as well. At the moment we have
no explanation for this, but the standard deviation of the residuals remaining in the Hipparcos data after
prewhitening with the first three frequencies is still higher (0.021 mag) than can be
expected from the errors on the measurements (0.017 mag). This could indicate that more (low
amplitude) frequencies are present, but they could not be detected in the individual data sets.
In the combined data set (Hipparcos, \citet{henry01} and Mercator data) a fourth frequency
at 0.8349\cd was detected.

We also observed a few times some extra brightening around light maximum
in this star. The standard deviation of the residuals in the Mercator data is about
0.015 mag in B1 and  0.011 mag in V. This is an indication of some additional
variability.

\subsection{\object{HD 2842} = HIP 2510}

In previous analyses of the Mercator data \citep{deridder04,cuypers06} only two
frequencies were found, but now the three frequencies, as given by \citet{henry05},
could be confirmed without any doubt.
The three-frequency solution is also clearly present in the Hipparcos data of this star.
The frequency given by \citet{koen02} for the Hipparcos data (1.78085\cd) is not present in any of the significant three-frequency solutions.

\citet{fekel03} observed no velocity changes in the 2 spectra they obtained, but
\citet{mathias04} did. The latter also mentioned possible line-profile variations
in the blue wing of the lines. 

\subsection{\object{HD 7169} = HIP 5674}

Three frequencies are easily recognized in the Mercator data.
Two frequencies are in common with the \citet{henry03} results, but
their second frequency (2.7640\cd) is probably an alias of the 
frequency 1.76031\cd\  found both in the Mercator and Hipparcos data.

This star has 136 Mercator measurements over the period 2001 to 2003, but was
unfortunately only once observed in 2004. Therefore, the errors on the results are larger than usual.

\subsection{\object{HD 23874} = HIP 17826} 

So far only one frequency (2.2565\cd) has been accepted for this star \citep{henry03}, 
since the second frequency found by \citet{handler99} in the Hipparcos data
could not be confirmed. However, more than one frequency is expected given the large 
residuals in all data sets. In the Mercator data a secondary frequency at 2.8883\cd\ or its
alias at 1.8856\cd\ is significant. A two-frequency search in the data of \citet{henry03}
gave also 2.2565 and 1.8849\cd\ as one of the most significant solutions. 
By combining the Henry and Fekel data, the Mercator data and the Hipparcos data, 
the two frequencies could be calculated more precisely.

\subsection{\object{HD 48271} = HIP 32263 = V553 Aur}

\citet{handler99} found two frequencies for this star in the Hipparcos data: 0.5244 and 0.8688\cd.
Only one frequency is given by \citet{henry03}: 0.9125\cd.
\citet{martin03} suggested the presence of four frequencies in the Hipparcos data.
Our analysis confirmd these four frequencies, although for the second frequency 
an alias is found to be more significant in the Mercator data 
and the fourth frequency is only marginally significant.
The combination of the data sets, with the Hipparcos data included,
solves the alias problems and indicates a stable four-frequency solution.

\subsection{\object{HD 62454} = HIP 37863 = DO Lyn}

For this star \citet{kaye99} found five frequencies: 1.60146, 1.43678,
1.73671,1.83372, 1.80753\cd\ in their multi-site campaign of 1997 and 1998 and they suggested one more near 2.9\cd. The first three have amplitudes larger than 0.01~mag in V and
are well identified. The others have smaller amplitudes. With our
multifrequency fit techniques, we found in the data of \citet{kaye99} the frequencies 1.60139, 1.43675, 1.73637,1.83373, 1.80793\cd, all very close to the given solution.

The most significant frequency in the Hipparcos data is 1.3329\cd . This frequency
is not related to any of the other frequencies, but this could be due to the odd time distribution of the Hipparcos data.

In the Mercator data of 2001 to 2004 the main frequency found in filters B and B1 is
1.5985\cd . This could be an alias of the first frequency given by \citet{kaye99}. 
In the other filters this frequency cannot unambiguously be
identified. A frequency at 1.2032\cd\ is present in all filters.

It seems impossible to connect the frequency solutions of the different campaigns for this star.
This could mean that the frequencies and/or the amplitudes in this star are not
stable over a longer period or, more likely, that a (different) combination of more frequencies 
is necessary to describe the data adequately.

The star is a double-lined spectroscopic binary \citep{kaye98} and orbital
parameters were given by \citet{kaye99} and \citet{mathias04}. Line
profile variations were observed in the primary star by \citet{mathias04}.

We do not expect a large influence of the secondary component on the
pulsational behaviour given the relatively long orbital period of 11.615~days, but currently we have no explanation for the variation of the frequency spectrum from season to season.

\subsection{\object{HD 69715} = HIP 40791}

HD 69715 was confirmed as a \gdor\ star by \citet{henry05}. The two frequencies found by them,
2.4566 and 2.4416\cd\ with error 0.0002\cd, are confirmed in the Mercator data, 2.45642 and 2.44120\cd\
with error 0.00007\cd. 
In the Hipparcos data a two-frequency solution with 2.0129 and 2.36462\cd\ is present, as already indicated by \cite{martin03},
but a well defined local minimum is found for 2.45630 and 2.44139\cd\ as well . Therefore, in the combination of all data sets,
highly accurate values for these frequencies could be found (errors less than 0.00001\cd).
The time span of the observations obtained by \citet{martin03} for this star was probably too short to separate these two frequencies. 
As a consequence, it is not surprising that other values for the frequencies were found, although one of their results, 
2.425\cd, is close to the two frequencies given here.
\citet{mathias04} observed no line-profile variations in the spectra, pointing to very low velocity amplitudes,
probably due to rotational line broadening ($\mathrm{v}\sin{i} = 145~km/s$).

\subsection{\object{HD 74504} = HIP 43062}

\begin{figure*}
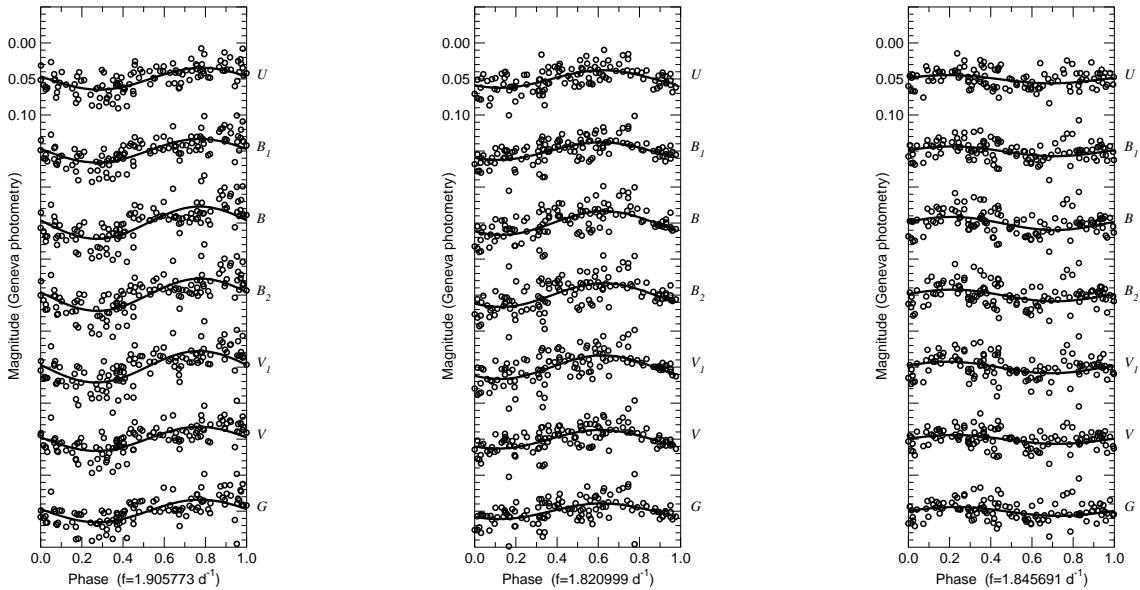

\begin{center}
\includegraphics[width=5.66cm]{74504_71.eps}
\includegraphics[width=5.66cm]{74504_72.eps}
\includegraphics[width=5.66cm]{74504_73.eps}
\end{center}
\caption[]{Phase diagrams of the seven-colour photometry of the newly discovered
\gdor\ star HD\,74504 for the indicated frequencies.}
\label{74504}
\end{figure*}

This is a star from the list by \citet{koen02} and it is the only star in our sample that is not yet catalogued as variable star.
It has all the properties of a \gdor\ star. The frequency in the Hipparcos data as listed by \citet{koen02} is 1.9057\cd.
This frequency ($1.90570\pm 0.00005$\cd) is found easily in the Mercator data, where at least one more frequency
could be identified ($1.8210 \pm 0.0001$\cd). Although this frequency could be an
alias, we prefer this value since a  two-frequency analysis of the Hipparcos data
gives $1.9058$ and $1.8212$\cd\ as the best two-frequency solution.
In the combined data set we found $1.905773$ and $1.820999$\cd.
The next frequency candidate here is $1.845691$\cd.
Phase diagrams constructed with the frequencies found are given in Figure \ref{74504}.

\subsection{\object{HD 86358} = HIP 48895 = HR 3936}

For this \gdor\ star, identified as such by \citet{handler99} and \citet{koen02}, four frequencies were known from the observations
published by \citet{fekel03}. In the Hipparcos data we identified three of those four frequencies.
In the  Mercator data the four frequencies (or some clear aliases) are significant as well, 
but an additional high amplitude variation with frequency 1.00286\cd\ (or an alias near 1.0014\cd) is present.
Its amplitude (0.022 mag in V, but smaller in other filters) is not comparable at all with the little extra power in the periodogram 
near 1\cd\ mentioned by \citet{fekel03} and seems large for a stellar oscillation frequency.
It is difficult to explain why a frequency corresponding to one cycle per sidereal day is so significant in this dataset
and only marginally or not at all significant in other datasets. It could be an artefact of e.g imperfect extinction corrections,
but in that case we would expect this frequency to be significant in other Mercator data sets as well. This is not the case.
After prewhitening with this frequency, it was possible to find a well-defined set of four frequencies in the Mercator data and in the combined data as well.
In a five-frequency fit, the variations associated with the frequency near 1\cd\ still have a significant amplitude. Therefore we calculated the associated amplitudes for this frequency as well. 

Indications of the binarity of this star were given by \citet{fekel03} and \citet{henry03}.
The spectra published by \citet{mathias04} clearly showed the binarity and indications of line-profile variations in the slow rotator.
\citet{griffin06} discussed the system extensively and gave revised orbital elements and mass estimates.
In principle, the frequency 1.00286\cd\ or 1.00139\cd\ found in the Mercator data could be a one day alias of a smaller frequency 
related to the orbital period of this system (33.7\,days), but this is unlikely in view of the estimated inclination of 45$\degr$ as given by \citet{griffin06}.

\subsection{\object{HD 100215} = HIP 56275}

This star is a well known \gdor\ star. The four frequencies given by \citet{henry03}, or daily aliases
are present in the Mercator data as well. The frequencies were refined by combining all data including the Hipparcos data, but some of the results can still be one year aliases of the true frequencies.
A candidate for a fifth frequency, present also in the combined data set, is 1.54354\cd.
The star is a binary system of which \citet{griffin06} gives high-precision orbital elements.

\subsection{\object{HD 105458} = HIP 59203}

\citet{henry01} found six frequencies in the data for this star.
We can easily confirm four of them (or their aliases) in the Mercator data.
The six-frequency solution is a local minimum as well and is also present in the combined V data.
Therefore, amplitude ratios for all frequencies could be estimated. In the Hipparcos data alone, the frequencies are difficult to find. 
There 0.7151\cd\ appears as a first frequency, although a frequency of 1.3210\cd\ is present as well.

\subsection{\object{HD 108100} = HIP 60571 = DD CVn}

The frequencies found in the Mercator data are compatible with the results of
\citet{breger97} and \citet{henry02}. The Hipparcos data are very noisy and could not give independent information.
\citet{breger97} indicated possible amplitude variations and this might be a reason for not finding the frequencies in a data set with a small number of observations. However, combining the Hipparcos data and the data of  \citet{henry02} with the Mercator data leads to a well-defined three-frequency solution. \citet{mathias04} report that this star is a binary with line-profile variations in the slow rotator.

\subsection{\object{HD 113867} = HIP 63951}

\citet{henry03} found five frequencies (0.8887, 2.0084, 1.7785, 2.8965 and 0.8841\cd) and they indicated that there could be some alias problems in their final solution. One of their frequencies is doubtful since it differs only by 0.0046\cd\ from the main frequency 
and this is at the limit of reliable extraction from an observational campaign of less than 250 days for this star.
In the Mercator data a similar solution could be found (0.8884, 1.0056,  1.7761, 1.8910, 0.8834\cd).
By combining the data of \citet{henry03} with the Mercator data and the Hipparcos observations, some ambiguities regarding day and year aliases could be resolved. A six-frequency solution (0.88838, 1.00420, 1.77757, 1.89378, 0.88681, 1.42221\cd) was obtained from this analysis.
The sixth frequency (or its alias at 2.42\cd) is also marginally present in Figure~18 of \citet{henry03}.
The frequency analysis for this star is not decisive. There is one period extremely close to 1 day. The third frequency nearly equals the sum of the first and fifth frequency. As a result not all alias problems nor ambiguities related to possible sum or double frequencies are completely resolved yet. Some indications of closely spaced frequencies near the main frequency are present. 
The star is very likely to be a binary, as suggested by \citet{henry03} and confirmed by \cite{mathias04}. It is not clear which component is the \gdor\ star. Since the components are estimated to have spectral type A9 and F1, both stars could be \gdor\ stars.

\subsection{\object{HD 167858} = HIP 89601 = HR 6844 = V2502 Oph}

There is no doubt that this binary star (orbital period: 4.48518 days \citep{fekel03}) is a multiperiodic variable. 
Two frequencies (near 0.765 and near 0.697\cd) are significant in all data sets.
The third frequency in the Mercator data is near the second frequency cited in \citet{aerts98}.
We identified five frequencies in the combined data, including the third frequency given by \citet{henry03}. 
One frequency is close to the $2-f$ alias of the second frequency and this may hamper the frequency analysis of individual ground-based data sets.
There is a lot of scatter in the light curves and some excess scatter at maximum brightness.

\subsection{\object{HD 175337} = HIP 92837}

One frequency is clearly present in all the data sets of this star (Hipparcos, \citet{henry05} and Mercator data).
The Hipparcos data have a few outliers that strongly influence the period search.
No secondary frequency could be identified unambiguously.

\subsection{\object{HD 195068} = HIP 100859 = HR 7828 = V2121 Cyg}

This star was sometimes observed on the same dates by \citet{henry05} and by the Mercator telescope team.
The three-frequency solution was easily confirmed in the combined data. 
A candidate for a fourth frequency occurs near 0.28 or its alias at 1.28\cd, between $f_1$ and $f_2$. 
The first frequency found by \citet{jankov06} in spectroscopic data ($1.61$\cd) is not significantly present in any of the photometric data sets.

\subsection{\object{HD 206043} = HIP 106897 = HR 8276 = NZ Peg}

For this star only three frequencies of the set of five found by \citet{henry01} were significant in the Mercator data.
However, a local minimum in the residuals exists for the five-frequency solution both in  the Mercator and in the combined data sets. 
One extra frequency was found in the combined data as well. Also for the six-frequency solution there was convergence to a local minimum in the residuals for the Mercator data.
Therefore, amplitudes could be calculated for all six frequencies in the Geneva filters. As expected the amplitudes of frequencies $f_4$, $f_5$ en $f_6$ are small.

\subsection{\object{HD 207223} = HIP 107558 = HR8330 =V372 Peg}

In the Mercator data the one year alias of the frequency given by \citet{aerts01} and \citet{guinan01} is found.
The Hipparcos data on its own are not very useful to resolve the ambiguity but in the combined data there is a slight preference for the frequency $0.38538$\cd.

\subsection{\object{HD 211699} = HIP 110163 = PR Peg}

Only three of the four frequencies (or of their aliases) found by \citet{henry07} are significant in the Mercator data. 
Their second frequency near 1.12\cd\ is not found, but is apparent in the combined data.
Furthermore, as already remarked by \citet{henry07}, their fourth frequency is equal to the difference between $f_1$ and $f_2$, or, $f_2 = f_1 +  f_4$. 
Adding a second harmonic to the fourth frequency significantly improves the fit. This adds to the puzzling state of this star as already reported by \citet{mathias04}, since line-profile variations were observed, but also changes of the equivalent widths of the line. Stellar activity is suggested as an explanation. This could mean that one of the frequencies might be related to rotation.

\subsection{\object{HD 218396} = HIP 114189 = HR 8799 = V342 Peg}

There is little doubt that this star has at least two frequencies.
The first two frequencies (1.9791 and 1.7368\cd) found in the multisite campaign presented in \citet{zerbi99} are present in the Mercator data as well ($1.9806$ and $1.7326$\cd). 
Since their multisite campaign was relatively short, there is agreement within the errors.
Combining the data with the Hipparcos measurements reveals another frequency,$0.76762$\cd, which could be an alias of the frequency $f_4=0.2479$\cd\ of \citet{zerbi99}.  

\subsection{\object{HD 221866} = HIP 116434}

It is known that for some time series of observations of multiperiodic variations, the technique of consecutive prewhitening will not yield the correct solution (see e.g. \citet{cuypers98}). The Hipparcos measurements of this star are another example of this phenomenon.
In the Hipparcos series a first  frequency appears near 0.90\cd, but only multifrequency fits lead to a solution compatible with the other sets of observations. In the final three-frequency solution, the frequency near 0.90\cd\ is not present anymore. When the Mercator and Hipparcos data were combined (the \citet{henry02} data were not used), very precise values for the frequencies could be found (errors less than 0.00001\cd).
The frequencies are not independent: in the Mercator data the third frequency is the mean of $f_1$ and $f_2$, in the data of \citet{henry02} and in the Hipparcos data the second frequency is the sum of the two others (within the errors).  Adding a harmonic to the third frequency of the Mercator data improves the fit. The two main frequencies are also in agreement with values published by \citet{kaye04}. Their spectroscopic observations indicate that HD~221866 is a spectroscopic binary with an orbital period of 135 days and a mass ratio of 1.11$\pm$0.03.

\section{Summary of the results}
The results of the frequency analysis of the Mercator data are summarized in Table \ref{resM}.
For the combined data the frequency solution can be found in Table \ref{res}. The errors on the frequencies are of the order of the last decimal given, typically $10^{-5}$ - $10^{-6}$\cd.

\begin{figure}
\centering
\resizebox{\hsize}{!}{\includegraphics[angle=270]{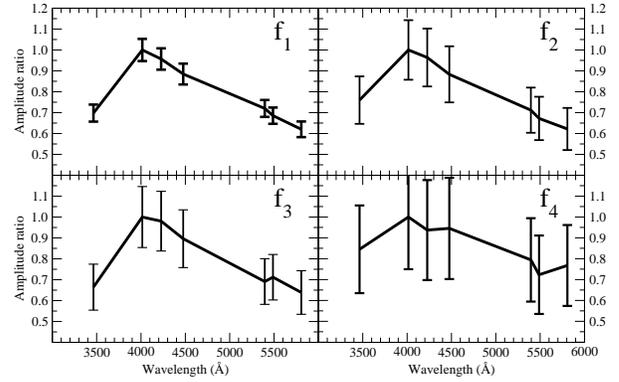}}
\caption{The ratios of the amplitudes in the Geneva colours (with reference to the highest amplitude) of the four frequencies found for the star HD~277}
\label{rat277}
\end{figure}

For all stars with a well-defined frequency solution in the combined data, the amplitudes in the different filters of the Geneva photometric systems were calculated. For each individual star, we give a summary of the results based on the Mercator data in Tables 5 to 25.  The errors on the frequencies (and the periods) were calculated with the expressions given by \citet{montgomery99} or \citet{breger99}, but only the largest value over the filters is given. In general the difference is less than a factor~of 2. The errors on the amplitudes were calculated with the standard deviations of the final residuals. As a consequence, the errors are the same for each frequency in a multifrequency solution. Since these values are larger than the formal values obtained from a harmonic least-squares fit, they can be considered as a reliable upper limit. The relative errors will increase and this is reflected in the increase of errors on the amplitude ratios from the dominant to the lower amplitude modes (see Figure \ref{rat277}).

The mean values of the magnitudes in each filter as obtained from the fit (with errors) are given in each Table as well.
The values were used for the calibration of effective temperature, gravity and a metallicity indicator according to \citet{kunzli97}, of which a summary is given in Table \ref{calib}.

\section{Conclusions}

Almost all frequencies of known \gdor\ stars presented in this paper are confirmed and one candidate (HD 74504) can now be considered as an additional member of this class. By combining the Mercator data 
with data available from the literature (including Hipparcos data), sets of frequencies were 
enlarged, confirmed and/or more precisely determined.
Many \gdor\ stars have frequencies that are very stable over a long period (1989-2004).
Only in a few cases did the frequencies found in the Hipparcos data (1989-1993)
deviate from the values found in more recent data. Differences are only a few hundredths
of a cycle per day in some cases and yearly aliasing could be the cause, but in a few other stars there was no agreement at all.
This may be caused by the coarse sampling of the Hipparcos data, but it could also be
an indication of some more hidden frequencies, complex beating patterns and/or 
changing frequencies. 

For the large majority of the \gdor\ stars the behaviour of the amplitude ratios as a function of the wavelength is very similar.
We will explore these observational results in future modelling of their oscillations, with the goal to check if we can constrain the input physics of the models for F-type stars.
Our study is the only consistent long-term multicolour photometric database of g-mode pulsations in an extended sample of \gdor\ stars so far, and it constitutes a useful reference for future interpretations of the oscillatory behaviour of such pulsators.

\begin{acknowledgements}
This research was made possible thanks to the financial support from the Fund for Scientific Research - Flanders (FWO), under projects G.0178.02 and G.0332.06
The Leuven authors additionally are supported by the Research Council of K.U.Leuven under grant GOA/2008/04.
The research leading to these results has received funding from the European
Research Council under the European Community's Seventh Framework Programme
(FP7/2007--2013)/ERC grant agreement n$^\circ$227224 (PROSPERITY).
We are indebted to our colleagues of the Geneva Observatory who reduced the \mercator\ data.
This research has made use of the NASA's Astrophysics Data System and the SIMBAD
astronomical database operated at the CDS in Strasbourg, France.
\end{acknowledgements}

\bibliographystyle{aa}
\bibliography{af}

\begin{table*}
\caption{Mercator data results. The spectral type is from literature. N is the number of observations used, T is the total time span in days. Frequencies not significant in all colour filters are indicated by a colon.} 
\begin{center}
\begin{tabular}{rrlrrllllll}
\hline\hline
    HD number & HIP number & Spectral  &    N  &  T(d) &  $f_1$(\cd)& $f_2$(\cd)&  $f_3$(\cd) & $f_4$(\cd) & $f_5$(\cd)&$f_6$(\cd)\\
              &            & type &      &  & & &  & & \\
\hline
\\
  277   &     623    &  F0    &  239 & 1148 & 1.11109 & 1.08130 & 1.38709 \\
 2842   &    2510    &  F0V   &  109 & 1149 & 1.71270 & 1.53729 & 1.67063 \\
 7169   &    5674    &  F2V   &   96 & 1149 & 1.8225  & 1.9245 \\
23874   &   17826    &  F0    &  127 & 1045 & 2.2567  & 1.8856 \\
48271   &   32263    &  F0    &   88 &  869 & 0.9123  &  0.8586 & 0.9314  & 1.7327 \\
62454   &   37863    &  F0    &  158 &  847 & 1.5985: \\
69715   &   40791    &  A5    &   64 &  805 & 2.45642 & 2.44120 \\
74504   &   43062    &  F0    &  122 &  794 & 1.90574 & 1.82101 & 1.84570 \\
86358   &   48895    &  F3V   &  100 &  805 & 1.1216  & 1.2930  & 1.1869  & 1.1423  & 1.00286 \\
100215  &   56275    &  Am    &  169 &  839 & 1.32191 & 1.42201 & 1.27911 & 1.6159:\\
105458  &   59203    &  F0III &  340 &  985 & 1.32084 & 1.25034 & 1.40903 & 0.94631 & 1.55279 & 1.09007\\
108100  &   60571    &  F2V   &  196 &  980 & 1.40132 & 1.32726 & 1.34071: \\
113867  &   63951    &  F0 &  163 &  946 & 0.88841 & 1.00553 & 1.89101 & 1.77613 & 0.88340 & 1.42185 \\
167858  &   89601    &  F2V   &   62 &  407 & 0.7650  & 0.6984& 1.0734 \\
175337  &   92837    &  F5    &   40 &  752 & 1.27109 \\
195068  &  100859    &  F5    &  189 &  766 & 1.25054 & 1.29843 & 0.96553 & 0.28517\\
206043  &  106897    &  F2V   &  158 & 1163 & 2.4324  & 2.3596  & 2.5243 \\
207223  &  107558    &  F3V   &  115 & 1163 & 0.38287 \\
211699  &  110163    &  F0    &   97 & 1149 & 0.9327  & 1.1963  & 1.1635 \\
218396  &  114189    &  A5V   &   95 & 1151 & 1.9806  & 1.7326  & 0.7676 \\
221866  &  116434    &  A3m &  135 & 1149 & 0.8384  & 0.8772  & 0.8572 \\
\hline
\end{tabular}
\label{resM}
\end{center}
\end{table*}

\begin{table*}
\caption{Frequencies obtained by combining data sets as described in the text. The definition of the colums is the same as in Table \ref{resM}. In the column "source" H means Hipparcos data, M Mercator data and L indicates that other data from literature were used as well. Values in bold indicate newly found frequencies, bold italics means an alias of a previously known frequency was used. }
\begin{center}
\begin{tabular}{rrlrrlllllll}
\hline\hline
    HD  & HIP & Spectral  &  Source &  N  &  T(d) &  $f_1$(\cd)& $f_2$(\cd)&  $f_3$(\cd) & $f_4$(\cd) & $f_5$(\cd)&$f_6$(\cd)\\
       number      &  number           & type &      &  & & &  & & \\
\hline
\\
  277   &     623    &  F0    & HLM & 864 & 5371 & 1.11101  & 1.08099  & 1.38704  & {\bf 0.83469} \\
 2842   &    2510    &  F0V   & HLM & 435 & 5377 & 1.71268  & 1.53725  & 1.66799 \\
 7169   &    5674    &  F2V   & HLM & 548 & 5384 & 1.82259  & 1.92502  & {\bf \em 1.76031} \\
23874   &   17826    &  F0    & HLM & 493 & 5385 & 2.25651  &  {\bf 1.88562} \\
48271   &   32263    &  F0    & HLM & 580 & 5109 &  {\bf \em 0.91236}  &   {\bf \em 0.87189}  &  {\bf \em 0.92314}  &  {\bf \em 1.75472}\\
62454   &   37863    &  F0    & HLM & 264 & 5084 & 1.60156  & 1.43631  & 1.73694  & 1.83413  & 1.81079 \\
69715   &   40791    &  A5    & HLM & 580 & 4816 & 2.456426 & 2.441492 \\
74504   &   43062    &  F0    & H~M & 213 & 5113 & 1.905773 &  {\bf 1.820999} &  {\bf 1.845691} \\
86358   &   48895    &  F3V   & HLM & 598 & 5169 & 1.289696 & 1.185229 & 1.121666 & 1.142454 &  {\bf 0.997646}: \\
100215  &   56275    &  Am    & HLM & 946 & 5213 & 1.32188  & 1.42207  & 1.27905  & 1.61480  &  {\bf 1.54354}\\
105458  &   59203    &  F0III & HLM &1043 & 5215 & 1.32083  & 1.25036  & 0.94632  & 1.40903  & 1.09021 & 1.55261\\
108100  &   60571    &  F2V   & HLM & 747 & 5192 & 1.32572  & 1.40136  & 1.36542 \\
113867  &   63951    &  F0 & HLM & 722 & 5180 & 0.88838  &  {\bf \em 1.00420}  &  {\bf \em 1.89379}  & 1.77757  & 1.42221\\
167858  &   89601    &  F2V   & HLM & 622 & 4532 & 0.76508  & 0.69845  & 1.30185  &  {\bf 1.60558}  &  {\bf 1.07551}\\
175337  &   92837    &  F5    & HLM & 372 & 4869 & 1.271134 \\
195068  &  100859    &  F5    & HLM & 610 & 4994 & 1.250511 & 1.298422 & 0.965416 &  {\bf 0.284863} \\
206043  &  106897    &  F2V   & HLM & 600 & 5366 & 2.35944  & 2.43242  & 2.52427  & 2.26551  & 2.59895 &  {\bf 2.46092} \\
207223  &  107558    &  F3V   & H~M & 204 & 5351 &  {\bf \em 0.385383} \\
211699  &  110163    &  F0    & HLM & 617 & 5472 & 0.93280  & 1.12646  & 1.16322  &  0.19377 \\
218396  &  114189    &  A5V   & H~M & 169 & 5394 & 1.980518 & 1.732528 &  {\bf \em 0.767562} \\
221866  &  116434    &  A3M & H~M & 304 & 5394 & 0.877197 & 0.838478 & 1.715664 \\
\hline
\end{tabular}
\label{res}
\end{center}
\end{table*}

\begin{table*}
\caption{Average magnitudes of the \gdor\ stars in the Geneva photometric system.}\label{colM}
\begin{center}
\begin{tabular}{rrrrrrrrr}
\hline\hline
    HD number & U & B1 & B & B2 & V1 & V & G & \\
    \hline\\
   277 &   9.2075 &  8.7915 &  7.8194 &  9.2134 &  9.0873 &  8.3666 &  9.4687 \\
  2842 &   8.7691 &  8.3682 &  7.4092 &  8.8170 &  8.7050 &  7.9858 &  9.0944 \\
  7169 &   8.1116 &  7.7034 &  6.7373 &  8.1385 &  8.0028 &  7.2810 &  8.3826 \\
 23874 &   9.0657 &  8.6702 &  7.6947 &  9.0866 &  8.9212 &  8.1980 &  9.2914 \\
 48271 &   8.2670 &  7.8668 &  6.9094 &  8.3160 &  8.1995 &  7.4807 &  8.5860 \\
 62454 &   7.9513 &  7.5519 &  6.5841 &  7.9808 &  7.8535 &  7.1336 &  8.2365 \\
 69715 &   7.9669 &  7.5578 &  6.6032 &  8.0114 &  7.8897 &  7.1708 &  8.2751 \\
 74504 &   9.6144 &  9.2039 &  8.2523 &  9.6614 &  9.5673 &  8.8519 &  9.9629 \\
 86358 &   7.2567 &  6.8881 &  5.9217 &  7.3235 &  7.1841 &  6.4625 &  7.5631 \\
100215 &   8.7886 &  8.3636 &  7.4040 &  8.8074 &  8.6996 &  7.9816 &  9.0866 \\
105458 &   8.5229 &  8.1213 &  7.1688 &  8.5793 &  8.4753 &  7.7584 &  8.8683 \\
108100 &   7.9628 &  7.5625 &  6.5931 &  7.9882 &  7.8371 &  7.1152 &  8.2101 \\
113867 &   7.6159 &  7.1823 &  6.2260 &  7.6367 &  7.5425 &  6.8253 &  7.9382 \\
167858 &   7.4093 &  6.9926 &  6.0277 &  7.4350 &  7.3337 &  6.6143 &  7.7259 \\
175337 &   8.2102 &  7.8103 &  6.8335 &  8.2302 &  8.1012 &  7.3784 &  8.4828 \\
195068 &   6.5117 &  6.0955 &  5.1233 &  6.5300 &  6.4220 &  5.6994 &  6.8119 \\
206043 &   6.5605 &  6.1317 &  5.1674 &  6.5809 &  6.4790 &  5.7572 &  6.8713 \\
207223 &   6.9961 &  6.6026 &  5.6300 &  7.0305 &  6.8966 &  6.1730 &  7.2762 \\
211699 &   9.9568 &  9.5352 &  8.5691 &  9.9679 &  9.8425 &  9.1226 & 10.2255 \\
218396 &   6.6524 &  6.2587 &  5.3186 &  6.7531 &  6.6774 &  5.9601 &  7.0853 \\
221866 &   8.2668 &  7.7669 &  6.8081 &  8.2177 &  8.1670 &  7.4528 &  8.5775 \\
\hline 
\end{tabular}
\end{center}
\end{table*}

\begin{table*}
\caption{Calibration of effective temperature, gravity and metallicity based on the Geneva colours of the \gdor\ stars. $\Delta$ is the formal calibration error.}\label{calib}
\begin{center}
\begin{tabular}{rrrrrrr} 
\hline\hline
HD number &   $T_{\mathrm{eff}}(\mathrm{K})$ &  $\Delta$ & $\log~\mathrm{g}$  & $\Delta$  &  $[M/H]$ & $\Delta$  \\
\hline
   277 & 6995 & 60 &  4.38 & 0.11 &  0.11 & 0.08  \\
  2842 & 7091 & 61 &  4.49 & 0.08 & -0.12 & 0.10  \\
  7169 & 6905 & 53 &  4.29 & 0.13 & -0.09 & 0.09  \\
 23874 & 6720 & 47 &  4.04 & 0.15 & -0.06 & 0.08  \\
 48271 & 7050 & 60 &  4.47 & 0.08 & -0.15 & 0.10  \\
 62454 & 6997 & 57 &  4.45 & 0.11 &  0.03 & 0.09  \\
 69715 & 6987 & 58 &  4.39 & 0.10 & -0.24 & 0.10  \\
 74504 & 7221 & 62 &  4.50 & 0.08 & -0.14 & 0.10  \\
 86358 & 6910 & 54 &  4.47 & 0.13 & -0.11 & 0.10  \\
100215 & 7109 & 59 &  4.40 & 0.07 & -0.04 & 0.09  \\
105458 & 7144 & 62 &  4.50 & 0.07 & -0.20 & 0.11  \\
108100 & 6809 & 50 &  4.17 & 0.15 & -0.08 & 0.09  \\
113867 & 7195 & 62 &  4.40 & 0.08 & -0.10 & 0.10  \\
167858 & 7177 & 61 &  4.46 & 0.07 &  0.01 & 0.09  \\
175337 & 6993 & 59 &  4.44 & 0.12 &  0.12 & 0.08  \\
195068 & 7133 & 61 &  4.46 & 0.08 &  0.07 & 0.09  \\
206043 & 7144 & 60 &  4.41 & 0.07 & -0.07 & 0.09  \\
207223 & 6952 & 54 &  4.43 & 0.12 &  0.01 & 0.09  \\
211699 & 6984 & 56 &  4.34 & 0.10 &  0.00 & 0.09  \\
218396 & 7355 & 67 &  4.57 & 0.07 & -0.71 & 0.18  \\
221866 & 7480 & 66 &  4.36 & 0.07 &  0.14 & 0.09  \\
\hline
\end{tabular}
\end{center}
\end{table*}

\end{document}